\documentclass[preprint,showpacs,aps,prd]{revtex4-1}
\usepackage{epsfig}
\usepackage{psfrag}
\usepackage{graphicx}
\usepackage{titlesec}
\usepackage{graphics}
\usepackage{amsmath}
\usepackage{color}
\usepackage[title]{appendix}

\begin{document}

%\begin{frontmatter}

\title{QED-vacuum response and Cherenkov radiated energy in non-linear and Lorentz-symmetry violating scenarios\footnote{This work is dedicated to the memory of our friend and collaborator Iv\'an Schmidt.}}

\author{Patricio Gaete} \email{patricio.gaete@usm.cl} 
\affiliation{Departamento de F\'{i}sica and Centro Cient\'{i}fico-Tecnol\'ogico de Valpara\'{i}so-CCTVal,
Universidad T\'{e}cnica Federico Santa Mar\'{i}a, Valpara\'{i}so, Chile}

\author{Jos\'e Abdalla Helay\"{e}l-Neto}\email{helayel@cbpf.br}
\affiliation{Centro Brasileiro de Pesquisas F\'{i}sicas (CBPF), Rio de Janeiro, RJ, Brasil} 

\date{\today}

\begin{abstract}
We investigate physical consequences of non-linear electrodynamic coupled to parameters that signal violation Lorentz-symmetry breaking (LSV). Our undertaking is done  by considering a general non-linear photonic Lagrangian which coupled to the Carroll-Field-Jackiw's model (CFJ). Our endeavor reveals how the (meta) material constitutive properties of the vacuum and wave propagation are affected by the interference of the LSV parameters LSV with the specific non-linear electrodynamic model under consideration. We also discuss the refractive indices for this new medium characterized by the coupling between non-linearities and the operators that carry the LSV message. Our results show that the QED-vacuum responds with birefringence and a dispersive propagation of waves. Subsequently, we consider the electromagnetic radiation produced by a moving charged particle interacting with this new medium. Our inspection illustrates that the emitted radiation reproduces the features of the Cherenkov effect for certain intensities of background magnetic fields . Finally, we compute the static potential profile within the framework of the gauge-invariant, but path-dependent, variables formalism. A logarithmic correction to the usual static Coulomb potential emerges driven by the LSV parameter and there also appear corrections due to the non-linearity; nevertheless, the logarithm behavior drops out whenever the LSV parameter is switched off.
\end{abstract}

%\pacs{14.70.-e, 12.60.Cn, 13.40.Gp}

\maketitle

\section{Introduction}

As is well known, Lorentz invariance is a fundamental ingredient of quantum field theory, which is an exact symmetry carried out by the Standard Model (SM) \cite{Kostelecky} of the interactions among the smallest building blocks of matter. This theory relies on Lorentz symmetry and provides a remarkably successful description of nowadays known phenomena. Interestingly enough, despite its experimental success, the observation of Lorentz symmetry violation (LSV) would indicate the existence of a new physics. In this form, the possibility that Lorentz and CPT symmetries be spontaneously broken at very fundamental scale, such as in string theories, has motivated a very intensive research activity. More precisely, the necessity of a new scenario has been proposed to overcome theoretical difficulties in the quantum gravity framework \cite{Amelino, Alfaro,Piran,Liberati}. It is worth recalling that theories with LSV are to be considered as effective theories and the analysis of its physical consequences at low energies may provide information and impose constraints on a more fundamental theory. It is to be specially noted that a suitable framework for testing the low-energy manifestations of LSV is the effective approach referred to as the Standard Model Extension (SME), where it is possible to realize spontaneous LSV.

On the other hand, quantum vacuum nonlinearities and its physical consequences such as vacuum birefringence and vacuum dichroism have been of great interest since its earliest days \cite{Euler,Adler,Tollis,Biswas, Seco, Michinel}. The work due to Euler and Heisenberg \cite{Euler} is the key example, who computed an effective nonlinear electromagnetic theory in vacuum emerging from the interaction of photons with virtual electron-positron pairs. Nevertheless, this amazing quantum characteristics of light has generated a growing interest on the experimental side \cite{Bamber,Burke,Pike}. For example, the PVLAS collaboration \cite{Valle}, and more recently the ATLAS collaboration has reported on the direct detection of the light-by-light scattering in LHC Pb-Pb collisions \cite{Aaboud, Enterria}. The advent of laser facilities has given rise to various proposals to probe quantum vacuum nonlinearities \cite{Battesti,Ataman}.

From the foregoing considerations, and given the ongoing experiments related to photon-photon interaction physics, we are encouraged to further examine how nonlinear electromagnetic effects couple to the parameters that signal LSV. To be more precise, we are interested in birefringence, Cherenkov radiation, as well as the static potential along the lines \cite{Logarithmic,EPL,Cherenkov23,GH23}. Seem from such a perspective, the present work is an expanded version of the study that was begun in \cite{Proceeding}. In this connection, it may be recalled that an electromagnetic theory coupled to parameters that signal LSV have been considered from different point of view \cite{Lehnert,Kaufhold,Borges}.

To carry out such studies we consider a non-linear electrodynamic coupled to a LSV term (actually, CFJ). We begin by computing the photonic dispersion relations to get how the parameters associated to the non-linearity interfere with the CFJ external vector in the 
(meta) material constitutive properties of the vacuum and group velocity. 
In fact, from the fourth-degree polynomial in the frequency can be shown to split into four categories: a purely Maxwellian term, a piece exclusively given by the non-linearity, a third part isolating the CFJ parameter and, finally, the desired contribution that couples LSV with the CFJ external vector. Furthermore, we compute the refractive indices illustrating 
that the new medium is birefringent as well as dispersive. 
Based on these results, let us discuss a more physically motivated aspect of the subject. To this end we consider the electromagnetic radiation produced by a moving charged particle interacting with this new quantum medium. To do this, we will work out the radiated energy following the usual approach of calculating the Poynting vector. Our analysis reveals the vital role played by vacuum electromagnetic nonlinearities with the CFJ term in triggering the radiated energy.

Our next objective is to study another aspect of the model under consideration, namely, the effect of the non-linearities with a CFJ term on a physical observable. To this end, we shall compute the static potential for this theory by using the gauge-invariant but path-dependent variables formalism. One important advantage of this approach is that it provides a physically alternative to the Wilson loop approach. As a result, we obtain a logarithmic correction to the usual static Coulomb potential. 

The paper is organized as follows: Section I introduces the model under consideration and provides the dispersion relations and the refractive indices. Section II discusses electromagnetic radiation, showing that the radiation obtained is like the one that happens in the Cherenkov effect. Section III is devoted to the calculation of the interaction energy for a pair of static probe charges within the framework of the gauge-invariant variables formalism. In particular, we shall be interested in the dependence of the parameters coming from the nonlinearities with the CFJ term. Finally, some final remarks are made in Section IV.

In our conventions the signature of the metric is ($+1,-1,-1,-1$).

\section{Model under consideration}

\subsection{General aspects}

We start off our considerations with a brief description of the model under consideration.The model is described by the following Lagrangian density: 
\begin{equation}
{\cal L} = {{\cal L}_{NL}}\left( {{\cal F};{\cal G}} \right) + {{\cal L}_{CFJ}},
 \label{NLCFJ05}
 \end{equation}
where the electromagnetic invariants (${\cal F}$;${\cal G}$) are given by ${\cal F} \equiv  - \frac{1}{4}{F_{\mu \nu }}{F^{\mu \nu }} = \frac{1}{2}\left( {\frac{{{{\bf E}^2}}}{{{c^2}}} - {{\bf B}^2}} \right)$ and ${\cal G} \equiv  - \frac{1}{4}{F_{\mu \nu }}{\tilde F^{\mu \nu }} = \frac{{\bf E}}{c} \cdot {\bf B}$. Here, ${{\cal L}_{NL}}$, describes the nonlinear part, whereas the term  Carroll, Field, Jackiw term, ${{\cal L}_{CFJ}}$, is given by
 \begin{equation}
{\cal L}_{CFJ} = \frac{1}{4}{\varepsilon ^{\mu \nu \kappa \lambda }}{v_\mu }{A_\nu }{F_{\kappa \lambda }}. \label{NLCFJ10}
 \end{equation}
 Here, ${v_\mu}$, is an arbitrary four-vector which selects a preferred direction in the space-time.
 
 Next, following our earlier procedure, we split the  $A_{\mu}$-field as the sum of a classical background $A_{B\mu}$, and a small quantum fluctuation, $a_{\mu}$. In such a case, 
the tensor $F_{\mu\nu}$ is given by
$F_{\mu\nu}=f_{\mu\nu}+F_{B\mu\nu}$, where $f^{\mu\nu}=\partial^{\mu}a^{\nu}-\partial^{\nu}a^{\mu}=\left( \, -e^{i} \, , \, -\epsilon^{ijk} \, b^{k} \, \right)$, while $F_{B}^{\;\,\mu\nu}=\partial^{\mu}A_{B}^{\;\,\nu}-\partial^{\nu}A_{B}^{\;\,\mu} =\left( \, -E_{B}^{\,\,i} \, , \, -\epsilon^{ijk} \, B_{B}^{\,\,k} \, \right)$. Accordingly, the corresponding equations of motion, up to quadratic terms in the fluctuations read
\begin{eqnarray}
\nabla  \cdot {\bf d} - {\bf v} \cdot {\bf b} = 0, \nonumber\\
\nabla  \times {\bf e} =  - \frac{{\partial {\bf b}}}{{\partial t}}, \nonumber\\
\nabla  \cdot {\bf b} = 0, \nonumber\\
\nabla  \times {\bf h} - {v^0}\, {\bf b} + {\bf v} \times {\bf e} = \frac{{\partial {\bf e}}}{{\partial t}},  \label{NLCFJ15}
\end{eqnarray}  
where the ${\bf d}$ and ${\bf h}$ fields are given by
\begin{equation}
 {\bf d} = \mathord{\buildrel{\lower3pt\hbox{$\scriptscriptstyle\leftrightarrow$}} 
\over \varepsilon }  \cdot {\bf e} + \mathord{\buildrel{\lower3pt\hbox{$\scriptscriptstyle\leftrightarrow$}} 
\over \zeta }  \cdot {\bf b}, \label{NLCFJ20}
\end{equation}
\begin{equation}
{\bf h} = {\mathord{\buildrel{\lower3pt\hbox{$\scriptscriptstyle\leftrightarrow$}} 
\over \mu } ^{ - 1}} \cdot {\bf b} + \mathord{\buildrel{\lower3pt\hbox{$\scriptscriptstyle\leftrightarrow$}} 
\over \eta }  \cdot {\bf e}, \label{NLCFJ25}
\end{equation}
in this case the tensors, $\mathord{\buildrel{\lower3pt\hbox{$\scriptscriptstyle\leftrightarrow$}} 
\over \zeta }$ and $\mathord{\buildrel{\lower3pt\hbox{$\scriptscriptstyle\leftrightarrow$}} 
\over \eta }$,  are given by    ${\zeta _{ij}} =  - c\frac{{{D_3}}}{{{C_1}}}{B_{Bi}}{B_{Bj}}$ and ${\eta _{ij}} = \frac{{{D_3}}}{{c\,{C_1}}}{B_{Bi}}{B_{Bj}}$.\\

It should be emphasized again that the vacuum electromagnetic properties are characterized by the following expressions for the vacuum permittivity and the vacuum permeability:
\begin{equation}
{\varepsilon _{ij}} \equiv {\delta _{ij}} + {\alpha _i}E_{Bj} + {\beta _i}B_{Bj},  \label{NLCFJ30}
\end{equation}
\begin{equation}
{\mu ^{ - 1}}_{ij} \equiv {\delta _{ij}} - B_{Bi}{\gamma _j} - E_{Bi}{\Delta _j}.  \label{NLCFJ35}
\end{equation}
Here we have simplified our notation by writing
\begin{equation}
\boldsymbol {\alpha}  \equiv \frac{1}{{{C_1}}}\left( {{D_1}\,{\bf E}_{B} + {D_3}\,{\bf B}_{B}} \right),\!\!\!\!\! \ \ \ \boldsymbol {\beta}  \equiv \frac{1}{{{C_1}}}\left( {{D_2}\, {\bf B}_{B} + {D_3} \,{\bf E}_{B}} \right),  \label{NLCFJ40}
\end{equation}
\begin{equation}
\boldsymbol {\gamma}  \equiv \frac{1}{{{C_1}}}\left( {{D_1}\, {\bf B}_{B} - {D_3} \,{\bf E}_{B}} \right),\!\!\!\! \ \ \ \boldsymbol {\Delta}  \equiv \frac{1}{{{C_1}}}\left( { - {D_3}\, {\bf B}_{B} + {D_2} \,{\bf E}_{B}} \right), \label{NLCFJ45}
\end{equation}
where
%%%%%%%%
\begin{eqnarray}
C_{1}=\left.\frac{\partial{\cal L}_{NL}}{\partial{\cal F}}\right|_{{\bf E}_{B},{\bf B}_{B}}
\, , \,
\left. C_{2}=\frac{\partial{\cal L}_{NL}}{\partial{\cal G}}\right|_{{\bf E}_{B},{\bf B}_{B}}
\, , \,
\nonumber \\
\left. D_{1}=\frac{\partial^2{\cal L}_{NL}}{\partial{\cal F}^2}\right|_{{\bf E}_{B},{\bf B}_{B}}
\, , \,
\left. D_{2}=\frac{\partial^2{\cal L}_{NL}}{\partial{\cal G}^2}\right|_{{\bf E}_{B},{\bf B}_{B}}
\, , \,
%\nonumber \\
\left. D_{3}=\frac{\partial^2{\cal L}_{NL}}{\partial{\cal F}\partial{\cal G}}\right|_{{\bf E}_{B},{\bf B}_{B}}. \label{NLCFJ50}
%\hspace{0.4cm}
\end{eqnarray}

%%%%%%%%%%%%%%%%%
Furthermore, restricting our considerations to the ${\bf E}_{B}=0$ case, we have that $D_{3}=0$. Making use of this result one encounters that equation (\ref{NLCFJ30}) becomes 
${\varepsilon _{ij}} = {\delta _{ij}} + \frac{{{D_2}}}{{{C_1}}}{B_{Bi}}{B_{Bj}}$, which have two eigenvalues $\varepsilon=1$ and $\varepsilon=1 + \frac{{{D_2}}}{{{C_1}}}{\bf B}_B^2$. Similarly, from equation 
(\ref{NLCFJ35}) we have 
$\mu _{ij}^{ - 1} = {\delta _{ij}} - \frac{{{D_1}}}{{{C_1}}}{B_{Bi}}{B_{Bj}}$. In this case the eigenvalues of $\mu _{ij}$ are given by $\mu  = 1$ and $\mu  = \frac{1}{{\left( {1 - \frac{{{D_1}}}{{{C_1}}}{\bf B}_B^2} \right)}}$.

%%%%%%%%%%%%%%%%%%% 
Mention should be made, at this point, to a further aspect related to vacuum electromagnetic properties. More specifically, we refer to:
 How do the parameters and external fields coming from the nonlinear (NL) sector couple to LSV parameters ($v^{\mu}$ for CFJ)? In what follows we will examine this question.

\subsection{Dispersion relations}

In order to adequately deal with this issue, one must inspect the dispersion relations (DRs) for an electromagnetic wave propagating in the external electromagnetic background; for the model under consideration, the matrix yielding the DRs reads as follows below:
\begin{equation}
\det {M_{ij}}\left( {w,{\bf k}; {\bf E}, {\bf B}_{B};{v^\mu }} \right) = 0, \label{NLCFJ55}
\end{equation}
where the $M$-matrix has the form
\begin{eqnarray}
{M_{ij}} &\equiv& {w^2}{\varepsilon _{ij}} + w{\zeta _{im}}{\varepsilon _{mnj}}{k_n} + {\varepsilon _{imn}}{\varepsilon _{klj}}{\mu ^{ - 1}}_{nk}{k_m}{k_l} \nonumber\\
 &+& w{\varepsilon _{imn}}{\eta _{nj}}{k_m} + i{v^0}{\varepsilon _{imj}}{k_m} - iw{\varepsilon _{imj}}{v_m}. \label{NLCFJ60}
 \end{eqnarray}

Next, we shall focus here on a single situation for the different nonlinear (NL) models:
${\bf E_{B}}= 0, {\bf B_{B}} \ne 0$. This decouples the coefficient $D_{3}$ from the problem and the $M$-matrix becomes:
\begin{eqnarray}
{M_{ij}} &=& \left[ {\left( {{w^2} - {{\bf k}^2}} \right) + \frac{{{D_1}}}{{{C_1}}}\left( {{\bf B}_B^2\,
{{\bf k}^2} - {{\left( {{{\bf B}_B} \cdot {\bf k}} \right)}^2}} \right)} \right]{\delta _{ij}} \nonumber\\
 &+&\left( {\frac{{{D_2}}}{{{C_1}}}{w^2} - \frac{{{D_1}}}{{{C_1}}}{{\bf k}^2}} \right){B_{Bi}}{B_{Bj}} + \left( {1 - \frac{{{D_1}}}{{{C_1}}}{\bf B}_B^2} \right){k_i}{k_j} \nonumber\\
 &+& \frac{{{D_1}}}{{{C_1}}}\left( {{\bf B}_{B} \cdot {\bf k}} \right)\left( {{k_i}{B_{Bj}} + {k_j}{B_{Bi}}} \right) + i{\varepsilon _{imj}}\left( {{v^0}{k_m} - w{v_m}} \right).  \label{NLCFJ65}
\end{eqnarray} 

Interestingly, the matrix from which the dispersion relations (DRs) follow can then be split into $4$ pieces:
\begin{equation}
DR= Max +NL+LS+ NL/LSV=0,   \label{NLCFJ70}
\end{equation}
where Max, NL, LSV and NL/LSV denote Maxwell, Nonlinear, Lorentz symmetry violation and the coupling between nonlinear and LSV, respectively.
These terms are given by:
\begin{equation}
Max = \left( {{w^2} - {{\bf k}^2}} \right)^{2},  \label{NLCFJ70a}
\end{equation}
\begin{eqnarray}
NL &=& \frac{{{D_2}}}{{{C_1}}}{{\bf B}_{B}^2}\,{w^4} - \frac{{{D_1}}}{{{C_1}}}{{\bf B}_{B}^2}\,{{\bf k}^4} + \left( {\frac{{{D_1}}}{{{C_1}}} - \frac{{{D_2}}}{{{C_1}}} + \frac{{{D_1}{D_2}}}{{C_1^2}}{{\bf B}_{B}^2}} \right){{\bf B}_{B}^2}\,{{\bf k}^2}\,{w^2},  \label{NLCFJ70b}
\end{eqnarray}
\begin{equation}
LSV =  - {\bf v^2}{w^2} + {{\bf v}^2}{{\bf k}^2} - {\left( {{\bf v} \cdot {\bf k}} \right)^2}. \label{NLCFJ70c}
\end{equation}

The coupled NL/LSV effects we are pursuing are cast in what follows:
\begin{eqnarray}
NL/LSV &\equiv&  - \frac{{{D_2}}}{{{C_1}}}{\left( {{{\bf B}_B} \cdot {\bf v}} \right)^2}{w^2} + \frac{{{D_1}}}{{{C_1}}}{\left( {{{\bf B}_B} \cdot {\bf v}} \right)^2}{\bf k^2} 
-\frac{{{D_1}}}{{{C_1}}}{\bf B}^2_{B}\left( {{{\bf v}^2}{{\bf k}^2} - {{\left( {{\bf v} \cdot {\bf k}} \right)}^2}} \right). \label{NLCFJ75}
\end{eqnarray}

It is worthy recalling, at this stage, that the coefficients $C_{1}$, $D_{1}$, $D_{2}$ specify the particular nonlinear electrodynamic model under consideration; where ${C_1}=\frac{{\partial {{\cal L}_{NL}}}}{{\partial {\cal F}}}$, ${D_1}=\frac{{{\partial ^2}{{\cal L}_{NL}}}}{{\partial {{\cal F}^2}}}$, ${D_2} = \frac{{{\partial ^2}{{\cal L}_{NL}}}}{{\partial {{\cal G}^2}}}$ and $D_{3}$ does not contribute whenever either ${\bf E}_{B}$ or ${\bf B}_{B}$ vanishes.

In this way, one encounters
\begin{equation}
m_\gamma ^2 = \frac{{{{\bf v}^2} - \frac{{{D_2}}}{{{C_1}}}{{\left( {{{\bf B}_B} \cdot {\bf v}} \right)}^2}}}{{1 + \frac{{{D_2}}}{{{C_1}}}{\bf B}_B^2}}.  \label{NLCFJ80}
\end{equation}
which represents, in the particle picture of the propagating mode, the photon (inertial) mass, which now depends on both sets of non-linear and LSV parameters, and also on the direction of the external magnetic field relative to the external CFJ space vector. Another aspect to be pointed out is that the shift induced by non-linearity on the CFJ photon effective mass, ${\mathrm{m}}_{\gamma}^{2} = {\left|{\mathrm{\bf v}}\right|}^{2}
$, vanishes whenever $D_2 = 0$, i.e., the shift is non-trivial only the non-linear Lagrangian density depends on ${\cal G}$  with a non-vanishing second
derivative, as given by the coefficient $D_2$. This is the case, for example, of the Born-Infeld, Euler-Lagrange and the Logarithmic non-linear models. Therefore, Lagrangians of the form ${\cal L}(\cal F)$ do not interfere on the photon effective mass induced by the LSV through the CFJ space vector.
%%%%%%%%

It is of interest also to notice that for logarithmic electrodynamics \cite{Logarithmic} 
\begin{equation}
{\cal L} =  - {\beta ^2}\ln \left[ {1 - \frac{{\cal F}}{{{\beta ^2}}} - \frac{{{{\cal G}^2}}}{{2{\beta ^4}}}} \right], \label{NLCFJ80-a}
\end{equation}
where ${\cal F} =  - \frac{1}{4}F_{\mu \nu }^2 = \frac{1}{2}\left( {{{\bf E}^2} - {{\bf B}^2}} \right)$ and ${\cal G} =  - \frac{1}{4}{F_{\mu \nu }}{\tilde F^{\mu \nu }} = {\bf E} \cdot {\bf B}$, the coefficients $C_1$, $D_1$ and $D_2$ are given by
\begin{eqnarray}
{C_1} = \frac{1}{{\left( {1 + \frac{{{\bf B}_B^2}}{{2{\beta ^2}}}} \right)}}, \,\,\,
{D_1} = \frac{1}{{{\beta ^2}}}\frac{1}{{{{\left( {1 + \frac{{{\bf B}_B^2}}{{2{\beta ^2}}}} \right)}^2}}}, \,\,
{D_2} = \frac{1}{{{\beta ^2}}}\frac{1}{{\left( {1 + \frac{{{\bf B}_B^2}}{{2{\beta ^2}}}} \right)}}. 
 \label{NLCFJ80-d}
 \end{eqnarray}

With the foregoing information, equation (\ref{NLCFJ80}) reduces to 
\begin{equation}
m_\gamma ^2 = \frac{{{{\bf v}^2} - \frac{{{{\left( {{{\bf B}_B} \cdot {\bf v}} \right)}^2}}}{{{\beta ^2}}}}}{{1 + \frac{{{\bf B}_B^2}}{{{\beta ^2}}}}}. \label{NLCFJ80-e}
\end{equation}

%%%%%%%
Proceeding further, the group velocity can be shown to result from the composition between the wave vector and the external CFJ space vector as written in the expression below:
\begin{equation}
{{\bf v}_g} = \frac{{\bf k}}{w}\,\frac{P}{Q} + \frac{{\bf v}}{w}\,\frac{R}{Q}.  \label{NLCFJ85}
\end{equation}
To get the last line we used
\begin{eqnarray}
P &\equiv&{w^2} - {{\bf k}^2} - \frac{1}{2}\left( {\frac{{{D_1}}}{{{C_1}}} - \frac{{{D_2}}}{{{C_1}}} + \frac{{{D_1}{D_2}}}{{C_1^2}}{\bf B}_B^{2}} \right){\bf B}_B^2\,{w^2} \nonumber\\
 &+& \frac{{{D_1}}}{{{C_1}}}{\bf B}_B^2\,{{\bf k}^2} - \frac{1}{2}{{\bf v}^2} + \frac{1}{2}\frac{{{D_1}}}{{{C_1}}}{\left( {{{\bf B}_B} \cdot {\bf v}} \right)^2} 
 - \frac{1}{2}\frac{{{D_1}}}{{{C_1}}}{\bf B}_B^2\,{{\bf v}^2}, \label{NLCFJ90}
\end{eqnarray}
\begin{eqnarray}
Q &\equiv& {w^2} - {{\bf k}^2} + \frac{1}{2}\left( {\frac{{{D_1}}}{{{C_1}}} - \frac{{{D_2}}}{{{C_1}}} + \frac{{{D_1}{D_2}}}{{C_1^2}}{\bf B}_B^2} \right){\bf B}_B^2\,{{\bf k}^2} \nonumber\\
&+&\frac{{{D_2}}}{{{C_1}}}{\bf B}_B^2\,{w^2} - \frac{1}{2}{{\bf v}^2} - \frac{1}{2}\frac{{{D_2}}}{{{C_1}}}{\left( {{{\bf B}_B} \cdot {\bf v}} \right)^2},  \label{NLCFJ95}
\end{eqnarray}
\begin{equation}
R \equiv \frac{1}{2}{\bf v} \cdot {\bf k} - \frac{1}{2}\frac{{{D_1}}}{{{C_1}}}{\bf B}_B^2\left( {{\bf v} \cdot {\bf k}} \right).  \label{NLCFJ100}
\end{equation}
%%%%%%

Now making use of equations  (\ref{NLCFJ80-d}), we readily find that
\begin{eqnarray}
P &=& {\omega ^2} - {{\bf k}^2} - \frac{{{\bf B}_B^4}}{{4{\beta ^4}\left( {1 + \frac{{{\bf B}_B^2}}{{2{\beta ^2}}}} \right)}}{\omega ^2} + \frac{{{\bf B}_B^2}}{{{\beta ^2}\left( {1 + \frac{{{\bf B}_B^2}}{{2{\beta ^2}}}} \right)}}{{\bf k}^2} \nonumber\\
&-& \frac{{{{\bf v}^2}}}{2} - \frac{1}{{2{\beta ^2}\left( {1 + \frac{{{\bf B}_B^2}}{{2{\beta ^2}}}} \right)}}{\left( {{{\bf B}_B} \times {\bf v}} \right)^2}, \label{NLCFJ100a}
\end{eqnarray}

\begin{eqnarray}
Q &=& {\omega ^2} - {{\bf k}^2} + \frac{{{\bf B}_B^4}}{{4{\beta ^2}\left( {1 + \frac{{{\bf B}_B^2}}{{2{\beta ^2}}}} \right)}}{{\bf k}^2} + \frac{{{\bf B}_B^2}}{{{\beta ^2}}}{\omega ^2} - \frac{{{{\bf v}^2}}}{2} 
- \frac{1}{{2{\beta ^2}}}{\left( {{{\bf B}_B} \cdot {\bf v}} \right)^2},  \label{NLCFJ100b}
\end{eqnarray}

\begin{equation}
R = \frac{1}{{2{\beta ^2}\left( {1 + \frac{{{\bf B}_B^2}}{{2{\beta ^2}}}} \right)}}\left( {{\bf k} \cdot {\bf v}} \right). \label{NLCFJ100c}
\end{equation}

The refractive indices, ${n_\bot}$ and ${n_\parallel}$, can also be found with the help of our DRs, given by det M = 0. Alternatively, however, these refractive indices can also be achieved via the equations of motion of the model under consideration. This will be considered in the next subsection.

\subsection{Refractive indices}

As already mentioned, our immediate objective is to calculate the refractive indices.
This can be readily accomplished by means of 
eqs.(\ref{NLCFJ20})-(\ref{NLCFJ45}). We mention in passing that, if $D_{3}=0$, we have $\mathord{\buildrel{\lower3pt\hbox{$\scriptscriptstyle\leftrightarrow$}} 
\over \zeta }  = \mathord{\buildrel{\lower3pt\hbox{$\scriptscriptstyle\leftrightarrow$}} 
\over \eta }  = 0$. Therefore, the electric permittivity 
($\varepsilon _{ij}$) and the inverse magnetic permeability 
($\left( {\mu}^{-1}  \right)_{ij}$) tensors of the vacuum, are also expressed as
\begin{equation}
{\varepsilon _{ij}} = {\delta _{ij}} + \frac{{{D_2}}}{{{C_1}}}{B_{Bi}}{B_{Bj}}, \ \ \
{\left( {{\mu ^{ - 1}}} \right)_{ij}} = {\delta _{ij}} - \frac{{{D_1}}}{{{C_1}}}{B_{Bi}}{B_{Bj}}, \label{refr05}
\end{equation}
with $d_i  = \varepsilon _{ij} e_j$ and $b_i  = \mu _{ij} h_j $.

Now, in order to describe the propagation of light in this new medium we will solve the source-free Maxwell equations (\ref{NLCFJ15}), taking eqs. (\ref{refr05}) into account.
 
According to the usual procedure we make a plane wave decomposition for the fields ${\bf e}$ and ${\bf b}$, so we write 
\begin{equation}
{\bf e}\left( {{\bf x},t} \right) = {\bf {\bar e}}\,{e^{ - i\left( {\omega t - {\bf k} \cdot {\bf x}} \right)}}, \ \ \
{\bf b}\left( {{\bf x},t} \right) = {\bf {\bar b}}\,{e^{ - i\left( {\omega t - {\bf k} \cdot {\bf x}} \right)}}.\label{refr10}
\end{equation}
This allows us to write the Maxwell equations as
\begin{equation}
{\bf k} \cdot \left( {\varepsilon {\bf {\bar e}}} \right) = {\bf v} \cdot {\bf {\bar b}}, \label{refr15-a}
\end{equation}
\begin{equation}
\omega \,{\bf {\bar b}} = {\bf k} \times {\bf {\bar e}}, \label{refr15-b}
\end{equation}
\begin{equation}
{\bf k} \cdot {\bf {\bar b}} = 0, \label{refr15-c}
\end{equation}
\begin{equation}
\omega\, {\bf {\bar d}} =  - {\bf k} \times \left( {{\mu ^{ - 1}}\,{\bf {\bar b}}} \right) + {\bf v} \times {\bf {\bar e}}.\label{refr15-d}
\end{equation}

Next, without restricting generality we take the $z$ axis as the direction of the magnetic field, ${\bf B_B}  = B_B\, \pmb {{\hat e}_3}$, and assuming that the light wave moves along the $x$ axis. In such a case, the tensors (\ref{refr05}) then are diagonal, that is, ${\varepsilon _{ij}} = diag\left( {1,1,1 + \frac{{{D_2}}}{{{C_1}}}{\bf B}_B^2} \right)$ and ${\left( {{\mu ^{ - 1}}} \right)_{ij}} = diag\left( {1,1,1 - \frac{{{D_1}}}{{{C_1}}}{\bf B}_B^2} \right)$. We further assume that ${\bf v} = v \,\pmb {{\hat e_3}}$.

So that the Maxwell equations become
\begin{equation}
\left( {\frac{{{k^2}}}{{{\omega ^2}}} - {\varepsilon _{22}}\,{\mu _{33}} + \frac{{{v^2}}}{{{\omega ^2}}}\frac{{{\mu _{33}}}}{{{\varepsilon _{11}}}}} \right){\bar e_2} = 0, \label{refr20-a}
\end{equation}
and
\begin{equation}
\left( {\frac{{{k^2}}}{{{w^2}}} - {\varepsilon _{33}}\,{\mu _{22}}} \right){\bar e_3} = 0. \label{refr20-b}
\end{equation} 

In view of these equations we have two different situations: First, if ${\bf {\bar e}}\ \bot \ {\bf B}_{B}$ (perpendicular polarization), from (\ref{refr20-b}) ${\bar e}_3=0$, and from (\ref{refr20-a}) we get $\frac{{{k^2}}}{{{\omega ^2}}} = {\varepsilon _{22}}\,{\mu _{33}} - \frac{{{v^2}}}{{{\omega ^2}}}\frac{{{\mu _{33}}}}{{{\varepsilon _{11}}}}$. In this way one encounters
\begin{equation}
{n_ \bot } = \sqrt {\left( {1 - \frac{{{{\bf v}^2}}}{{{\omega ^2}}}} \right)} \frac{1}{{\sqrt {\left( {1 - \frac{{{D_1}}}{{{C_1}}}{\bf B}_B^2} \right)} }}.  \label{refr25}
\end{equation}

Second, if ${\bf {\bar e}}\ || \ {\bf B}_B$ (parallel polarization), from (\ref{refr20-a}) ${\bar e}_2=0$, and from (\ref{refr20-b}) we get $\frac{{{k^2}}}{{{w^2}}} = {\varepsilon _{33}}\,{\mu _{22}}$. We may therefore write 
\begin{equation}
{n_\parallel } = \sqrt {1 + \frac{{{D_2}}}{{{C_1}}}{\bf B}_B^2}. \label{refr30} 
\end{equation}

This implies that the preceding electromagnetic vacuum acts like a birefringent medium with two refractive indices determined by the polarization of the incoming electromagnetic waves. What is however quite peculiar in this case is the fact that, according to equation (\ref{refr25}), the new medium is also a dispersive one, driven by the Lorentz breaking vector. While from equation (\ref{refr30}) it is clear that the refractive index is due to the non-linearities of the model under consideration, but with the particularity that it is actually controlled by the coefficient $D_2$ , in the same way as it happens with the shift on the effective photon mass, as discussed in the paragraph below eq. (\ref{NLCFJ80}). 

Let us, for later convenience, introduce the following notation ${\Gamma _B} \equiv \frac{{{\bf B}_B^2}}{{2{\beta ^2}}} > 0$. From this expression it follows that for logarithmic electrodynamics $\frac{{{D_1}}}{{{C_1}}} = \frac{1}{{{\beta ^2}}}\frac{1}{{\left( {1 + {\Gamma _B}} \right)}}$ and $\frac{{{D_2}}}{{{C_1}}} = \frac{1}{{{\beta ^2}}}$. We thus find that equations (\ref{refr25}) and (\ref{refr30}) become
\begin{equation}
n_ \bot ^2 = \left( {1 + {\Gamma _B}} \right)\left( {1 - \frac{{{{\bf v}^2}}}{{{\omega ^2}}}} \right), \label{refr30a}
\end{equation}
and
\begin{equation}
n_\parallel ^2 = 1 + \frac{{{\bf B}_B^2}}{{{\beta ^2}}}.\label{refr30b}
\end{equation}

\section{Cherenkov radiation}

Inspired by these observations, the purpose of this Section is to further elaborate on the physical content of the model under consideration. Our aim here is to examine the problem of obtaining the electromagnetic radiation produced by a moving charged particle interacting with this new medium along the lines of \cite{GH23}. To this end, we consider the Maxwell equations 
\begin{eqnarray}
\nabla  \cdot {\bf e} - \frac{{{\bf v} \cdot {\bf b}}}{\varepsilon } = \frac{{4\pi }}{\varepsilon }{\rho _{ext}}, \nonumber\\
\nabla  \cdot {\bf b} = 0, \nonumber\\
\nabla  \times {\bf e} =  - \frac{1}{c}\frac{{\partial {\bf b}}}{{\partial t}}, \nonumber\\
\nabla  \times {\bf b} + \mu \,{\bf v} \times {\bf e} = \frac{{4\pi \mu }}{c}{{\bf J}_{ext}} + \frac{{\varepsilon \mu }}{c}\frac{{\partial {\bf e}}}{{\partial t}}. \label{rad-05}
\end{eqnarray}
Here, ${\rho _{ext}}$ and ${{\bf J}_{ext}}$ denote the external charge and current densities, which are given by: ${\rho _{ext}}\left( {t,{\bf x}} \right) = Q\,\delta \left( x \right)\delta \left( y \right)\delta \left( {z - {\bar u}t} \right)$ and ${\bf J}_{ext}\left( {t,{\bf x}} \right) = Q\,{\bar u}\,\delta \left( x \right)\delta \left( y \right)\delta \left( {z - {\bar u}t} \right){\hat {\bf e}_z}$. We mention in passing that, for simplicity, we are considering the $z$ axis as the direction of the moving charged particle. In addition ${\bar u}$ is the velocity of the particle. 

Following our earlier procedure \cite{GH23}, we first observe that the previous equations can be brought to the form
\begin{widetext}
\begin{subequations}
\begin{eqnarray}
\left( { {\nabla ^2} - \frac{1}{{{c^{ \prime 2}}}}\frac{{{\partial ^2}}}{{\partial {t^2}}}} \right){\bf e} + \frac{\mu }{c}{\bf v} \times \frac{{\partial {\bf e}}}{{\partial t}} = \frac{{4\pi \mu }}{{{c^2}}}\frac{{\partial {{\bf J}_{ext}}}}{{\partial t}} + \frac{{4\pi }}{\varepsilon }{\bf \nabla} {\rho _{ext}} + \frac{1}{\varepsilon }\nabla \left( {{\bf v}  \cdot {\bf b}} \right), 
\label{rad-10a} \\
 \left( {{\nabla ^2} - \frac{1}{{{c^{ \prime 2}}}}\frac{{{\partial ^2}}}{{\partial {t^2}}}} \right){\bf b} + \frac{\mu }{\varepsilon }{\bf v}\left( {{\bf v} \cdot {\bf b}} \right) = \frac{{4\pi \mu }}{\varepsilon }{\bf v}{\rho _{ext}} - \frac{{4\pi }}{c}\nabla  \times {{\bf J}_{ext}} - \mu \left( {{\bf v} \cdot \nabla } \right){\bf e}.\label{rad-10b}
 \end{eqnarray}
 \end{subequations}
 \end{widetext}
In the above we have defined $\frac{1}{{c}^{\prime{2}}}\equiv\frac{\varepsilon {\mu}}{{c}^{2}}$. Furthermore, it is worth highlighting that these equations in a general setup are rather complicated, however, for an interesting physical case we are able to obtain exact solutions.
It is for this reason that in this work we will only consider terms of linear order in ${\bf v}$ $\left({{\mathrm{\cal O}}\left({\mathrm{\bf v}}\right)}\right)$. This assumption arises from current estimates in the literature \cite{Russell} on the parameters associated to the Lorentz- symmetry violating, the parameter $v \equiv |{\bf v}|$ must be upper-bounded ${v} < {10}^{{-}{42}}$ GeV.

As was explained in \cite{GH23}, to solve the above equations we first Fourier transform to momentum space via
\begin{equation}
F(t,{\bf x}) = \int {\frac{{dw{d^3}{\bf k}}}{{{{\left( {2\pi } \right)}^4}}}} {e^{ - iwt + i\,{\bf k} \cdot {\bf x}}}F\left( {w,{\bf k}} \right), \label{rad-15}
\end{equation}
where $F$ stands for the electric and magnetic fields. 

%%%%%%%%%
Making use of (\ref{rad-15}), we thus find that equation (\ref{rad-10a}) becomes
\begin{eqnarray}
{\cal O}\,{\bf e} - \frac{{i\mu }}{c}\omega \left( {{\bf v} \times {\bf e}} \right) &=& \frac{{i4\pi }}{\varepsilon }{\rho _{ext}}{\bf k} + \frac{i}{\varepsilon }\left( {{\bf v} \cdot {\bf b}} \right){\bf k} 
- \frac{{i4\pi \mu }}{{{c^2}}}\omega \,{\bf J}_{ext}, \label{rad-15a}
\end{eqnarray}
where we have defined ${\cal O} \equiv  - {{\bf k}^2} + \frac{{{\omega ^2}}}{{{c^{ \prime 2}}}}$.                                               
We can rewrite the previous equation as 
\begin{eqnarray}
\underbrace {\left( {{\cal O}{\delta _{ij}} + i{b_0}{\varepsilon _{ijk}}{v_k}} \right)}_{{M_{ij}}}{e_j} &=& i\frac{{4\pi }}{\varepsilon }{\rho _{ext}}{k_i} + \frac{i}{\varepsilon }\left( {{\bf v} \cdot {\bf b}} \right){k_i} 
- i\frac{{4\pi \mu }}{{{c^2}}}\omega \,{\left( {{{\bf J}_{ext}}} \right)_i}, \label{rad-15b}
\end{eqnarray}
where ${b_0} = \frac{{\mu \omega }}{c}$. It may be observed here that the inverse matrix of $M_{ij} = {{\cal O}{\delta _{ij}} + i{b_0}{\varepsilon _{ijk}}{v_k}}$, is given by $M_{jm}^{ - 1} = \frac{{\cal O}}{{\left( {{{\cal O}^2} - b_0^2{{\bf v}^2}} \right)}}{\delta _{jm}} - \frac{{b_0^2}}{{{\cal O}\left( {{{\cal O}^2} - b_0^2{{\bf v}^2}} \right)}}{v_j}{v_m} - \frac{{i{b_0}}}{{\left( {{{\cal O}^2} - b_0^2{{\bf v}^2}} \right)}}{\varepsilon _{jml}}{v_l}$.

We thus find that
\begin{eqnarray}
{\bf e}\left( {\omega ,{\bf k}} \right) &=& i\frac{{4\pi }}{\varepsilon }\frac{\rho_{ext} }{\cal O}{\bf k} + \frac{i}{\varepsilon }\frac{{\left( {{\bf v} \cdot {\bf b}} \right)}}{\cal O}{\bf k} - i\frac{{4\pi \mu }}{{{c^2}}}\frac{\omega }{\cal O}{{\bf J}_{ext}} \nonumber\\
&-& \frac{{4\pi \mu \omega \rho_{ext} }}{\varepsilon }\frac{{\left( {{\bf k} \times {\bf v}} \right)}}{{{{\cal O}^2}}} 
+ \frac{{4\pi {\mu ^2}{\omega ^2}}}{{{c^2}}}\frac{{\left( {{\bf J}_{ext} \times {\bf v}} \right)}}{{{{\cal O}^2}}}. 
\label{rad-15c}
\end{eqnarray} 

Furthermore, also using equation (\ref{rad-15}), we see that equation (\ref{rad-10b}) becomes
\begin{eqnarray}
{\cal O}\,{\bf b} + \frac{\mu }{\varepsilon }\left( {{\bf v}\cdot {\bf b}} \right){\bf v} &=& \frac{{4\pi \mu }}{\varepsilon }\,{\bf v} \,{\rho _{ext}} - i\frac{{4\pi \mu }}{c}\,{\bf k} \times {{\bf J}_{ext}} 
- i\mu \left( {{\bf k} \cdot {\bf v}} \right){\bf e}.  \label{rad-20}
\end{eqnarray}
 
Inserting the expression (\ref{rad-15c}) in equation (\ref{rad-20}), we obtain, to order ${\cal O}({\bf v})$, that the magnetic field takes the form
\begin{eqnarray}
\underbrace {\left( {{\cal O}\,{\delta _{ij}} + {d_0}\,{v_i}{v_j}} \right)}_{{G_{ij}}}{b_j} &=& \frac{{4\pi \mu }}{\varepsilon }{\rho _{ext}}{v_i} - i\frac{{4\pi \mu }}{c}{\left( {{\bf k} \times {{\bf J}_{ext}}} \right)_i} \nonumber\\
&+& \frac{{4\pi \mu {\rho _{ext}}}}{\varepsilon }\frac{{\left( {{\bf k} \cdot {\bf v}} \right)}}{\cal O}{k_i} 
- \frac{{4\pi {\mu ^2}}}{{{c^2}}}\frac{\omega }{\cal O}\left( {{\bf k} \cdot {\bf v}} \right){\left( {{{\bf J}_{ext}}} \right)_i}, 
\label{rad-25a}
\end{eqnarray}
in this last line we used ${d_0} = \frac{\mu }{\varepsilon }$. Let us also observe here that
the inverse matrix of ${G_{ij}} = {\cal O}\,{\delta _{ij}} + {d_0}\,{v_i}\,{v_j}$, can be written as $G_{jm}^{ - 1} = \alpha \, {\delta _{jm}} + \beta \,{v_j}{v_m}$. In this case, the coefficients $\alpha$ and $\beta$ are given by $\alpha  = \frac{1}{{\det G}}\,{\cal O}\left( {{\cal O} + {d_0}\,{{\bf v}^2}} \right)$, $\beta  =  - {d_0}\,\frac{{\cal O}}{{\det G}}$, whereas $\det G = {{\cal O}^2}\left( {{\cal O} + {d_0}\,{{\bf v}^2}} \right)$.\\

We may accordingly rewrite equation (\ref{rad-25a}), to order ${\cal O}({\bf v})$, in the form 
\begin{eqnarray}
{\bf b}\left( {\omega ,{\bf k}} \right) &=& \frac{{4\pi \mu }}{\varepsilon }{\bf v}\frac{{{\rho _{ext}}}}{\cal O} - i\frac{{4\pi \mu }}{c}\frac{{\left( {{\bf k} \times {{\bf J}_{ext}}} \right)}}{\cal O} \nonumber\\
&+& \frac{{4\pi \mu }}{\varepsilon }{\rho _{ext}}\frac{{\left( {{\bf k} \cdot {\bf v}} \right)}}{{{{\cal O}^2}}}{\bf k} 
- \frac{{4\pi {\mu ^2}}}{{{c^2}}}\omega \frac{{\left( {{\bf k} \cdot {\bf v}} \right)}}{{{{\cal O}^2}}}{{\bf J}_{ext}}. 
\label{rad-25b}
\end{eqnarray}

Making use of this last equation, we can write equation (\ref{rad-15c}) as
\begin{eqnarray}
{\bf e}\left( {\omega ,{\bf k}} \right) &=& i\frac{{4\pi {\rho _{ext}}}}{\varepsilon }\frac{\bf k}{\cal O} - i\frac{{4\pi \mu }}{{{c^2}}}\frac{\omega }{\cal O}{{\bf J}_{ext}} + \frac{{4\pi \mu }}{{c\varepsilon }}\frac{{\left[ {{\bf v} \cdot \left( {{\bf k} \times {\bf J}} \right)} \right]}}{{{{\cal O}^2}}}{\bf k} \nonumber\\
&-&\frac{{4\pi \mu \omega {\rho _{ext}}}}{{\varepsilon c}}\frac{{\left( {{\bf k} \times {\bf v}} \right)}}{{{{\cal O}^2}}}
+ \frac{{4\pi {\mu ^2}{\omega ^2}}}{{{c^3}}}\frac{{\left( {{\bf J} \times {\bf v}} \right)}}{{{{\cal O}^2}}}. \label{rad-25c}
\end{eqnarray}

In summary then, equations (\ref{rad-25b}) and (\ref{rad-25c}) form the basis of our subsequent analysis.

Next, following our earlier line of argument \cite{GH23}, we are in a position to calculate ${\bf b}\left( {w,{\bf x}} \right)$ and ${\bf e}\left( {w,{\bf x}} \right)$. We also recall that, ${\bf b}\left( {w,{\bf x}} \right)$ is given by
\begin{equation} 
{\bf b}\left( {w,{\bf x}} \right) = \int {\frac{{{d^3}{\bf k}}}{{{{\left( {2\pi } \right)}^3}}}} \; {e^{i{\bf k} \cdot {\bf x}}}\;{\bf b}\left( {w,{\bf k}} \right). \label{rad-30}
\end{equation}

Again, as in  \cite{GH23}, by making use of the axial symmetry of the problem under consideration and using cylindrical coordinates, we find that the magnetic field (\ref{rad-30}) can be brought to the form
%\begin{widetext}
\begin{eqnarray}
{\bf b}\left( {w,{\bf x}} \right) &=& \frac{{\mu Q}}{{\pi \varepsilon \bar u}}\,{v_z}{e^{i{{\omega z} \mathord{\left/
 {\vphantom {{\omega z} {\bar u}}} \right.
 \kern-\nulldelimiterspace} {\bar u}}}}\int_0^\infty  {d{k_T}\,{k_T}} \int_0^{2\pi } {d\alpha }\, \frac{{{e^{i{k_T}{x_T}\cos \alpha }}}}{{{\cal O}{|_{{k_z} = {\omega  \mathord{\left/
 {\vphantom {\omega  {\bar u}}} \right.
 \kern-\nulldelimiterspace} {\bar u}}}}}}\, \pmb {{{\hat e}_z}} \nonumber\\
 &-& i\frac{{\mu Q}}{{\pi c}}{e^{i{{\omega z} \mathord{\left/
 {\vphantom {{\omega z} {\bar u}}} \right.
 \kern-\nulldelimiterspace} {\bar u}}}}\int_0^\infty  {d{k_T}\,{k_T}} \int_0^{2\pi } {d\alpha } \,\frac{{{e^{i{k_T}{x_T}\cos \alpha }}}}{{{\cal O}{|_{{k_z} = {\omega  \mathord{\left/
 {\vphantom {\omega  {\bar u}}} \right.
 \kern-\nulldelimiterspace} {\bar u}}}}}}\left[ {{k_T}sen\alpha\, \pmb{\hat \rho}  - {k_T}\cos \alpha\, \pmb {\hat \phi }} \right] \nonumber\\
 &+& \frac{{\mu Q\omega }}{{\pi \varepsilon {{\bar u}^2}}}{v_z}{e^{i\omega {z \mathord{\left/
 {\vphantom {z {\bar u}}} \right.
 \kern-\nulldelimiterspace} {\bar u}}}}\int_0^\infty  {d{k_T}} {k_T}\int_0^{2\pi } {d\alpha } \left\{ {\frac{{{k_T}\left[ {\cos \alpha\, \pmb{\hat \rho}  + sen\alpha\, \pmb{\hat \phi} } \right] + {\raise0.5ex\hbox{$\scriptstyle \omega $}
\kern-0.1em/\kern-0.15em
\lower0.25ex\hbox{$\scriptstyle {\bar u}$}}  {\pmb{{\hat e}_z}}}}{{{{\cal O}^2}{|_{{k_z} = {\omega  \mathord{\left/
 {\vphantom {\omega  {\bar u}}} \right.
 \kern-\nulldelimiterspace} {\bar u}}}}}}} \right\}{e^{i{k_T}{x_T}\cos \alpha }} \nonumber\\
&-&\frac{{{\mu ^2}Q\omega }}{{\pi {c^2}}}{v_z}\frac{\omega }{{\bar u}}{e^{i\omega {z \mathord{\left/
 {\vphantom {z {\bar u}}} \right.
 \kern-\nulldelimiterspace} {\bar u}}}}\int_0^\infty  {d{k_T}} {k_T}\int_0^{2\pi } {d\alpha \frac{{{e^{i{k_T}{x_T}\cos \alpha }}}}{{{{\cal O}^2}{|_{{k_z} = {\omega  \mathord{\left/
 {\vphantom {\omega  {\bar u}}} \right.
 \kern-\nulldelimiterspace} {\bar u}}}}}}}   \pmb {{{\hat e}_z}}, \label{rad-35}
\end{eqnarray}
%\end{widetext}
where $\pmb{\hat \rho}$ and $\pmb{\hat \phi}$ are unit vectors normal and tangencial to the cylindrical surface, respectively. While $\pmb {{{\hat e}_z}}$ is a unit vector along the $z$ direction. We also note that the subscript $T$ in $k_{T}$ indicates transversal to the $z$ direction. It should be further noted that ${\left. {\cal O} \right|_{{k_z} = {w \mathord{\left/
 {\vphantom {w {\bar u}}} \right.
 \kern-\nulldelimiterspace} {\bar u}}}} = {w^2}\left( {\frac{1}{{{c^{\prime 2}}}} - \frac{1}{{{{\bar u}^2}}}} \right) - {\bf k}_T^2$.

Let us also recall here that
$\int_0^{2\pi } {d\theta } {e^{ix\cos \theta }}\sin \theta  = 0, 
\int_0^{2\pi } {d\theta } {e^{ix\cos \theta }}\cos \theta  = 2\pi i{J_1}\left( x \right), 
 \int_0^{2\pi } {d\theta } {e^{ix\cos \theta }} = 2\pi {J_0}\left( x \right)$,
where ${{J_0}\left( x \right)}$ and ${{J_1}\left( x \right)}$ are Bessel functions of the first kind. 

With the aid of the previous integrals, we find that  equation (\ref{rad-35}) becomes
%\begin{widetext}
\begin{eqnarray}
{\bf b}\left( {w,{\bf x}} \right) &=& \frac{{2 \mu Q}}{{ \varepsilon \bar u}}{v_z}{e^{i{{\omega z} \mathord{\left/
 {\vphantom {{\omega z} {\bar u}}} \right.
 \kern-\nulldelimiterspace} {\bar u}}}}\int_0^\infty  {d{k_T}\,{k_T}} \frac{{{J_0}\left( {{k_T}{x_T}} \right)}}{{{\cal O}{|_{{k_z} = {\omega  \mathord{\left/
 {\vphantom {\omega  {\bar u}}} \right.
 \kern-\nulldelimiterspace} {\bar u}}}}}} \,\pmb {{\hat e}_z} 
-\!\! \frac{{2\mu Q}}{c}{e^{i{{\omega z} \mathord{\left/
 {\vphantom {{\omega z} {\bar u}}} \right.
 \kern-\nulldelimiterspace} {\bar u}}}}\int_0^\infty  {d{k_T}\,k_T^2} \,\frac{{{J_1}\left( {{k_T}{x_T}} \right)}}{{{\cal O}{|_{{k_z} = {\omega  \mathord{\left/
 {\vphantom {\omega  {\bar u}}} \right.
 \kern-\nulldelimiterspace} {\bar u}}}}}}\,\pmb {\hat \phi} \nonumber\\
&+&\!\! i\frac{{2\mu Q\omega }}{{\varepsilon {{\bar u}^2}}}{v_z}{e^{i\omega {z \mathord{\left/
 {\vphantom {z {\bar u}}} \right.
 \kern-\nulldelimiterspace} {\bar u}}}}\int_0^\infty  {d{k_T}k_T^2} \frac{{{J_1}\left( {{k_T}{x_T}} \right)}}{{{{\cal O}^2}{|_{{k_z} = {\omega  \mathord{\left/
 {\vphantom {\omega  {\bar u}}} \right.
 \kern-\nulldelimiterspace} {\bar u}}}}}} \pmb {\hat \rho}   
+\!\! \frac{{2\mu Q\omega }}{{\varepsilon {{\bar u}^2}}}{v_z}{e^{i\omega {z \mathord{\left/
 {\vphantom {z {\bar u}}} \right.
 \kern-\nulldelimiterspace} {\bar u}}}}\frac{\omega }{{\bar u}}\int_0^\infty  {d{k_T}{k_T}} \frac{{{J_0}\left( {{k_T}{x_T}} \right)}}{{{{\cal O}^2}{|_{{k_z} = {\omega  \mathord{\left/
 {\vphantom {\omega  {\bar u}}} \right.
 \kern-\nulldelimiterspace} {\bar u}}}}}}\pmb{\hat e_z} \nonumber\\
&-&\!\! \frac{{2{\mu ^2}Q\omega }}{{{c^2}}}{v_z}{e^{i\omega {z \mathord{\left/
 {\vphantom {z {\bar u}}} \right.
 \kern-\nulldelimiterspace} {\bar u}}}}\frac{\omega }{{\bar u}}\int_0^\infty  {d{k_T}{k_T}} \frac{{{J_0}\left( {{k_T}{x_T}} \right)}}{{{{\cal O}^2}{|_{{k_z} = {\omega  \mathord{\left/
 {\vphantom {\omega  {\bar u}}} \right.
 \kern-\nulldelimiterspace} {\bar u}}}}}}\pmb{\hat e_z}. 
\label{rad-40}
\end{eqnarray}
%\end{widetext}

Then, by integrating over $k_T$ and performing further manipulations, we finally obtain that equation (\ref{rad-40}) may be rewritten as
%\begin{widetext}
\begin{eqnarray}
{\bf b}\left( {w,{\bf x}} \right) &=& - \frac{{\pi \mu Q\omega }}{{2\varepsilon {{\bar u}^2}}}{v_z}{e^{i\omega {z \mathord{\left/
 {\vphantom {z {\bar u}}} \right.
 \kern-\nulldelimiterspace} {\bar u}}}}\rho H_0^{\left( 1 \right)}\left( {\sigma \rho } \right)\pmb {\hat \rho} 
 + i \frac{{\pi \mu Q}}{c}{e^{i{{\omega z} \mathord{\left/
 {\vphantom {{\omega z} {\bar u}}} \right.
 \kern-\nulldelimiterspace} {\bar u}}}} \sigma H_1^{\left( 1 \right)}\left( {\sigma ,\rho } \right) \pmb {\hat \phi} \nonumber\\
&-& i\frac{{\pi \mu Q}}{{\varepsilon \bar u}}{v_z}{e^{i\omega {z \mathord{\left/
 {\vphantom {z {\bar u}}} \right.
 \kern-\nulldelimiterspace} {\bar u}}}}\left[ {H_0^{\left( 1 \right)}\left( {\sigma \rho } \right) - \frac{{\sigma \rho }}{2}H_1^{\left( 1 \right)}\left( {\sigma \rho } \right)} \right] \pmb{\hat e_z}, 
 \label{rad-45}
\end{eqnarray}
%\end{widetext}
where ${\mathit{\sigma}}^{2} = {\mathit{\omega}}^{2}\left({\frac{1}{{c}^{\prime {2}}}{-}\frac{1}{{v}^{2}}}\right)$. In the above we have used ${x_T} = \rho$ (in cylindrical coordinates). We also recall that
${H}_{0}^{(1)}(x)$ and ${H}_{1}^{(1)}(x)$ are Hankel functions of the first kind.

Before we proceed further, we call attention to the fact that the above expression for the magnetic field  (\ref{rad-45}) satisfies Maxwell's equation $
\nabla\cdot{\mathrm{\bf b}} = 0$. In this context we also point out that, in a previous work \cite{GH23}, we have shown in detail how the Cherenkov angle emerges in our framework. More precisely, it was shown that wave planes are radiated if the velocity of the charge is larger the velocity of light in the medium. 

We come now to the calculation of the electric field. By proceeding in the same way as before, the electric field turns out to be
%\begin{widetext}
\begin{eqnarray}
{\bf e}\left( {\omega ,{\bf x}} \right) &=&- \frac{{2Q}}{{\varepsilon \bar u}}{e^{iz{\omega  \mathord{\left/
 {\vphantom {\omega  {\bar u}}} \right.
 \kern-\nulldelimiterspace} {\bar u}}}}\int_0^\infty  {d{k_T}} k_T^2\frac{{{J_1}\left( {{k_T}{x_T}} \right)}}{{{\cal O}{|_{{k_z} = {\omega  \mathord{\left/
 {\vphantom {\omega  {\bar u}}} \right.
 \kern-\nulldelimiterspace} {\bar u}}}}}}\pmb{\hat \rho} 
+ \frac{{2\,i\,Q\,\mu\, {v_z}}}{{\varepsilon\, c\, \bar u}}\,\omega\, {e^{iz{\omega  \mathord{\left/
 {\vphantom {\omega  {\bar u}}} \right.
 \kern-\nulldelimiterspace} {\bar u}}}}\int_0^\infty  {d{k_T}\,k_T^2} \frac{{{J_1}\left( {{k_T}{x_T}} \right)}}{{{{\cal O}^2}{|_{{k_z} = {\omega  \mathord{\left/
 {\vphantom {\omega  {\bar u}}} \right.
 \kern-\nulldelimiterspace} {\bar u}}}}}}\, \pmb {\hat \phi} \nonumber\\
 &+& i\frac{{2Q}}{{\varepsilon \bar u}}{e^{iz{\omega  \mathord{\left/
 {\vphantom {\omega  {\bar u}}} \right.
 \kern-\nulldelimiterspace} {\bar u}}}}\frac{\omega }{{\bar u}}\int_0^\infty  {d{k_T}} {k_T}\frac{{{J_0}\left( {{k_T}{x_T}} \right)}}{{{\cal O}{|_{{k_z} = {\omega  \mathord{\left/
 {\vphantom {\omega  {\bar u}}} \right.
 \kern-\nulldelimiterspace} {\bar u}}}}}}\pmb{\hat e_z} - i\frac{{2\mu Q\omega }}{{{c^2}}}{e^{iz{\omega  \mathord{\left/
 {\vphantom {\omega  {\bar u}}} \right.
 \kern-\nulldelimiterspace} {\bar u}}}}\int_0^\infty  {d{k_T}} {k_T}\frac{{{J_0}\left( {{k_T}{x_T}} \right)}}{{{\cal O}{|_{{k_z} = {\omega  \mathord{\left/
 {\vphantom {\omega  {\bar u}}} \right.
 \kern-\nulldelimiterspace} {\bar u}}}}}}\pmb{\hat e_z}. \nonumber\\
 \label{rad-50}
 \end{eqnarray}
%\end{widetext}

Again, essentially following the same steps as in the case of the magnetic field, we find that the electric field becomes
%\begin{widetext}
\begin{eqnarray}
{\bf e}\left( {\omega ,{\bf x}} \right) &=& i\frac{{\pi Q}}{{\varepsilon \bar u}}{e^{iz{\omega  \mathord{\left/
 {\vphantom {\omega  {\bar u}}} \right.
 \kern-\nulldelimiterspace} {\bar u}}}}\sigma H_1^{\left( 1 \right)}\left( {\sigma \rho } \right)\pmb{\hat \rho}
- \frac{{\pi \mu Q\omega }}{{2c\varepsilon \bar u}}{v_z}{e^{iz{\omega  \mathord{\left/
 {\vphantom {\omega  {\bar u}}} \right.
 \kern-\nulldelimiterspace} {\bar u}}}}\rho H_0^{\left( 1 \right)}\left( {\sigma \rho } \right)\pmb{\hat \phi} \nonumber\\
&-& \frac{{\pi Q}}{\varepsilon }{e^{iz{\omega  \mathord{\left/
 {\vphantom {\omega  {\bar u}}} \right.
 \kern-\nulldelimiterspace} {\bar u}}}}\omega \left( {\frac{1}{{{c^{\prime 2}}}} - \frac{1}{{{{\bar u}^2}}}} \right)H_0^{\left( 1 \right)}\left( {\sigma \rho } \right)\pmb {\hat e_z}. 
 \label{rad-55}
\end{eqnarray}
%\end{widetext}

Equations  (\ref{rad-45}) and (\ref{rad-55}) will then be exploited to compute the radiated energy for the model under consideration.

Before going ahead, it is appropriate to observe that, as was explained in \cite{GH23}, the above development leads to the Cherenkov angle. In particular, we have showed that wave planes are radiated if the velocity of the charge is larger the velocity of light in the medium. We skip all the technical details and refer to \cite{GH23} for them. In view of this situation, we pass now to the calculation of the radiated energy.

We start by recalling that the density of power carried out by the radiation fields across the surface $S$ bounding the volume $V$ is given by the real part of the Poynting vector (time averaged value)
\begin{equation}
 {\bf S} = \frac{c}{{2\pi }}{\mathop{\rm Re}\nolimits} \left( {{\bf e} \times {{\bf b}^ * }} \right).                        \label{rad-60}
\end{equation}

Using this expression we can compute the power radiated through the surface $S$ \cite{Das} by means of
 \begin{equation}
{\cal E} = \int_{ - \infty }^\infty \! {dt} \int\limits_S {d{\bf a} \cdot {\bf S}}. \label{rad-65}
\end{equation}
One can now further observe that we shall consider a cylinder of length $h$ and radius $\rho$.
Hence the total energy radiated through the surface of the cylinder can be reduced to the form  
\begin{equation}
{\mathcal{E}} = \frac{c}{{2\pi }}\left( {2\pi \rho h} \right)\int_0^\infty  {d\omega }\, {S_\rho }\,\Theta \left( {\bar u - \frac{c}{n}} \right), \label{rad-70}
\end{equation}
where it may be recalled that the presence of the step function means that Cherenkov radiation will be emitted if and only if the velocity of the particle exceeds the velocity of light in the medium (${\bar u} > \frac{c}{n}$).

Using equations (\ref{rad-45}) and (\ref{rad-55}) in the radiation zone, we can express the power radiated per unit length (\ref{rad-70}) as
\begin{equation}
W \equiv \frac{\cal E}{h} = \frac{{2\,\pi\, {Q^2}}}{{{c^2}}}\frac{\mu }{\varepsilon }\int_0^\infty  {d\omega } \,{n^2}\left( \omega  \right)\omega \left( {1 - \frac{{{c^2}}}{{{n^2}\left( \omega  \right){{\bar u}^2}}}} \right). \label{rad-75}
\end{equation}
It is useful to point out that the above expression is analogous to that obtained in the Cherenkov radiation theory, except for an extra $n^2(w)$ in the integrand of equation (\ref{rad-75}).                 
 
Let us also mention here that the new vacuum is analogous to the general case of non-linear electrodynamics \cite{Cherenkov23,GH23}, where the polarizable medium is the QED vacuum, made up of virtual electrons and positrons. In this manner  the produced Cherenkov radiation corresponds to the energy re-emitted by the excited virtual particles,
hence we may set an upper bound for the integration over $w$ in (\ref{rad-75}).         
 As was explained in \cite{Cherenkov23,GH23}, we argue that the frequency, $\frac{2mc^2}{\hbar}$, acts as a cutoff frequency for the re-emitted photons, which then corresponds to the energy for a pair creation. 

With this, equation (\ref{rad-75}) can be written as  
\begin{equation}
W = \frac{{2\,\pi \,{Q^2}}}{{{c^2}}}\frac{\mu }{\varepsilon }\int_0^\Omega  {d\omega } \,{n^2}\left( \omega  \right)\omega \left( {1 - \frac{{{c^2}}}{{{n^2}\left( \omega  \right){{\bar u}^2}}}} \right),  \label{rad-80}
\end{equation}
where  $\Omega  = \frac{2mc^2}{\hbar}$.

From this expression, we can now proceed to estimate the energy lost per unit distance travelled for a charged particle. We first observe, by considering $\hbar  = c = 1$, that $\bar u > {1 \mathord{\left/{\vphantom {1 n_ \bot}} \right.\kern-\nulldelimiterspace} n_\bot}$. Second, from $n_ \bot ^2 > 1$ and equation (\ref{refr30a}), we find that ${\omega _{low}} > \sqrt {\frac{{1 + {\Gamma _B}}}{{{\Gamma _B}}}} \left| {\bf v} \right|$. Furthermore, according to Ref. \cite{Russell} we have $\left| {\bf v} \right| \sim {10^{ - 23}}MeV$ and using frequencies in the visible, that is, $\omega  \approx {10^{ - 6}}MeV$, we have $\frac{{{{\bf v}^2}}}{{{\omega ^2}}} \sim {10^{ - 34}} \ll 1$. We accordingly deduce that $n_ \bot ^2 \simeq \left( {1 + {\Gamma _B}} \right)$. Also, we readily verify that
$\frac{\mu }{\varepsilon } = \frac{{1 + {\Gamma _B}}}{{\left( {1 - {\Gamma _B}} \right)\left( {1 + 2{\Gamma _B}} \right)}}$.

We thus find that equation (\ref{rad-80}) reads
\begin{eqnarray}
W &=& 2\pi {e^2}\frac{\mu }{\varepsilon }{n^2}\left( {1 - \frac{1}{{{n^2}{{\bar u}^2}}}} \right)\int_{{\omega _{low}}}^{^\Omega } {d\omega }\, \omega \nonumber\\
&=&2\pi {e^2}\frac{{1 + {\Gamma _B}}}{{\left( {1 - {\Gamma _B}} \right)\left( {1 + 2{\Gamma _B}} \right)}}\left( {1 + {\Gamma _B}} \right) 
\left( {1 - \frac{1}{{\left( {1 + {\Gamma _B}} \right){{\bar u}^2}}}} \right)\frac{1}{2}\left( {4{m^2} - \frac{{\left( {1 + {\Gamma _B}} \right)}}{{{\Gamma _B}}}{{\left| {\bf v} \right|}^2}} \right). \nonumber\\
\end{eqnarray}
It is of interest also to notice that ${\bar u^2} > \frac{1}{{1 + {\Gamma _B}}}$, $4{m^2} \cong {\left( {1.02} \right)^2}Me{V^2}$, $|{\bf v}{|^2} \cong {10^{ - 46}}Me{V^2}$, ${e^2} = \frac{1}{{137}}$ and ${\Gamma _B} = \frac{{{\bf B}_B^2}}{{2{\beta ^2}}} = \frac{{{\bf B}_B^2}}{{35.28Me{V^4}}}$. As we can see $\Gamma_B$ depends on the background magnetic field ${\bf B}_B$. It is for this reason that below we will consider three different background magnetic fields:
\begin{itemize}
\item  $|{{\bf B}_B}| = 10\,\,T$  (Tokamaks)
\item $|{{\bf B}_B}| = 10^{9}\,\,T$  (Magnetars)
\item $|{{\bf B}_B}| = 10^{14}\,\,T$  (RHIC)
\end{itemize}

Firstly, for the Tokamaks case, we find that ${\Gamma _B} = \frac{{{{10}^2}{T^2}}}{{35.28Me{V^4}}} \cong {10^{ - 18}}$, leading to ${n_ \bot } \cong 1$. Obviously, in this case there is no Cherenkov radiation.

For the magnetars case, we have ${\Gamma _B} = \frac{{{{10}^{18}}{T^2}}}{{35.28Me{V^4}}} = \frac{1}{{73.17}}$ leading to ${n_ \bot } \cong {\left( {1 + {\Gamma _B}} \right)^{{1 \mathord{\left/
 {\vphantom {1 2}} \right.\kern-\nulldelimiterspace} 2}}} \cong 1.005$. In such a case, from $\bar u > \frac{1}{{{n_ \bot }}}$, we obtain $\bar u > {\left( {1 + 0.005} \right)^{ - 1}}$. Consequently, $\bar u > 0.995$ (ultra relativistic). Besides, $\frac{\mu }{\varepsilon } \sim 1$, $\left( {1 + {\Gamma _B}} \right) \cong 1.01    $, ${\bar u^2} \cong 0.99$, $4{m^2} \cong {\left( {1.02} \right)^2}Me{V^2}$ and $\frac{{\left( {1 + {\Gamma _B}} \right)}}{{{\Gamma _B}}}|v{|^2} \cong 0$. Thus, we finally obtain $W \cong  - 36.578\frac{{MeV}}{{\mu m}}$.
 
Finally, for the RHIC case, we have ${\Gamma _B} \cong 1.37 \times {10^8}$, ${n_ \bot } \cong {10^4}$, 
$\frac{\mu }{\varepsilon } \cong  - 0.5 \times {10^{ - 8}}$ and ${\bar u^2} > {10^{ - 8}}$. By considering, ${\bar u^2} = {10^{ - 6}}$, which then yields $W =  - 8.2 \times {10^4}\frac{{MeV}}{{\mu m}}$.

\section{Interaction energy}

We turn our attention to the calculation of the interaction energy between static point-like sources for our model described by equation (\ref{NLCFJ05}), by using the gauge-invariant but path-dependent variables formalism. However, before proceeding with the determination of the interaction energy, we first note that to study quantum properties of the electromagnetic field in the presence of external electric and magnetic fields, we should split the  $A_{\mu}$-field as the sum of a classical background $A_{B\mu}$, and a small quantum fluctuation, $a_{\mu}$, in the same way as we done in Sec. I.

In this manner, following our earlier procedure, the nonlinear part of Lagrangian (\ref{NLCFJ05}) up to quadratic terms in the fluctuations, is also expressed as 
 \begin{eqnarray}
{\cal L}_{NL}^{(2)} &=& -\frac{1}{4} \, C_{1} \, f_{\mu\nu}^{\, 2}
-\frac{1}{4} \, C_{2} \, f_{\mu\nu} \, \widetilde{f}^{\mu\nu}
-\frac{1}{2} \, G_{B\mu\nu} \, f^{\mu\nu} 
+ \,\frac{1}{8} \, Q_{B\mu\nu\kappa\lambda} \, f^{\mu\nu} \, f^{\kappa\lambda} \; , \label{ener05}
\end{eqnarray}
where $f^{\mu\nu}$ has been defined previously in Sec. I, and $\widetilde{f}^{\mu\nu}=\epsilon^{\mu\nu\alpha\beta}f_{\alpha\beta}/2$. We further recall that $G_{B\mu\nu}=C_{1} \, F_{B\mu\nu}+ C_{2} \, \widetilde{F}_{B\mu\nu}$ and
$Q_{B\mu\nu\kappa\lambda}=D_{1} \, F_{B\mu\nu}F_{B\kappa\lambda}
+D_{2} \, \widetilde{F}_{B\mu\nu}\widetilde{F}_{B\kappa\lambda}
+D_{3} \, F_{B\mu\nu}\widetilde{F}_{B\kappa\lambda}
+ D_{3} \, \widetilde{F}_{B\mu\nu} F_{B\kappa\lambda}$ are tensors
that depend on the components of the electric and magnetic background fields.
 Whereas, the coefficients $C_{1}$, $C_{2}$, $D_{1}$, $D_{2}$ and $D_{3}$ of this expansion are given by eq. (\ref{NLCFJ50}).

We accordingly rewrite the density Lagrangian (\ref{NLCFJ05}), up to quadratic terms in the fluctuations, in the form
\begin{eqnarray}
{\cal L} &=& -\frac{1}{4} \, C_{1} \, f_{\mu\nu}^{\, 2}
-\frac{1}{4} \, C_{2} \, f_{\mu\nu} \, \widetilde{f}^{\mu\nu}
-\frac{1}{2} \, G_{B\mu\nu} \, f^{\mu\nu} 
+\!\!\!\frac{1}{8} \, Q_{B\mu\nu\kappa\lambda} \, f^{\mu\nu} \, f^{\kappa\lambda} \; 
+ \frac{1}{4}{\varepsilon ^{\mu \nu \kappa \lambda }}{v_\mu }{a_\nu }{f_{\kappa \lambda }} - a_{0}J^{0}, \nonumber\\ 
\label{ener10}
\end{eqnarray}
where we have introduced an external current $J^{0}$.

This effective theory provide us with a suitable starting point to study the interaction energy. In order to obtain the corresponding Hamiltonian, the canonical quantization of this theory from the Hamiltonian analysis point of view is straightforward and follows closely that of Ref. \cite{EPL,Logarithmic,Neves23}. The canonical momenta read
\begin{eqnarray}
{\Pi ^\mu } &=&  - {C_1}{f^{0\mu }} - {C_2}{\tilde f^{0\mu }} - G_B^{0\mu } + \frac{1}{2}Q_B^{0\mu \kappa \rho }{f_{\kappa \rho }} 
+ \frac{1}{2}{\varepsilon ^{\kappa \rho 0\mu }}{v_{\kappa \rho }}. \label{ener15}
\end{eqnarray}
This yields the usual primary constraint ${\Pi ^0} = 0$. While the remaining nonzero momenta are given by
\begin{equation}
{\Pi ^i} =  - {C_1}{f^{0i}} - {C_2}{\tilde f^{0i}} - G_B^{0i} + \frac{1}{2}Q_B^{0i\kappa \rho } + \frac{1}{2}{\varepsilon ^{\kappa \rho 0i}}{v_\kappa }{a_\rho }. \label{ener20}
\end{equation}

The canonical Hamiltonian is now obtained in the usual way 
\begin{eqnarray}
{H_C} &=& \int {{d^3}x} \left\{ { - {a_0}\left( {{\partial _i}{\Pi ^i} - \frac{1}{2}{\varepsilon ^{ijk}}{v_i}{\partial _j}{a_k} - {J^0}} \right)} \right\} \nonumber\\
 &+& \int {{d^3}x} \left\{ {\frac{1}{2}\left( {{C_1}{\delta ^{ki}} + Q_B^{k0i0}} \right){f_{i0}} - \frac{1}{2}{C_2}{{\tilde f}^{k0}}{f_{k0}}} \right\} \nonumber\\
&+& \int {{d^3}x} \left\{ {\frac{1}{4}Q_B^{k0ij}{f_{ij}}{f_{k0}} + \frac{1}{4}{C_1}{f_{ij}}{f^{ij}} + \frac{1}{2}{G_{Bij}}{f^{ij}}} \right\} 
+ \int {{d^3}x} \left\{ {\frac{1}{4}{C_2}{f_{ij}}{{\tilde f}^{ij}}} \right\}. \label{ener25}
\end{eqnarray}

Next, in what follows we will consider the case of a pure magnetic background, that is, ${{\bf E}_B} = 0$ and ${{\bf B}_B} \ne 0$. We thus find that
\begin{eqnarray}
{H_C} &=& \int {{d^3}x} \left\{ { - {a_0}\left( {{\partial _i}{\Pi ^i} + \frac{1}{2}{\bf v} \cdot {\bf b} - {J^0}} \right) + \frac{1}{2}{C_1}{{\bf e}^2}} \right\}   \nonumber\\
 &+& \int {{d^3}x} \left\{ {\frac{1}{2}{D_2}\,{{\left( {{{\bf B}_B} \cdot {\bf e}} \right)}^2} + \frac{1}{2}{C_1}{{\bf b}^2} + {C_1}{{\bf B}_B} \cdot {\bf b}} \right\}. 
\label{ener30}
\end{eqnarray}

Time conserving the primary constraint ${\Pi}_0$ yields the secondary constraint ${\Gamma _1} \equiv {\partial _i}{\Pi ^i} + \frac{1}{2}{\bf v} \cdot {\bf b}-J^{0} = 0$ (Gauss's law). Therefore, in this case there are two constraints in all, which are first class. Now we recall that the extended Hamiltonian, which generates the time evolution of the dynamical variables, is then written as $H = H_C  + \int {d^3 x} \left( {u_0(x) \Pi_0(x)  + u_1(x) \Gamma _1(x) } \right)$. Here $u_o(x)$ and $u_1(x)$ are arbitrary Lagrange multipliers reflecting the gauge invariance of the theory. Considering that $\Pi^0=0$ always and ${\dot {a_0}}\left( x \right) = \left[ {{a_0}\left( x \right),H} \right] = {u_0}\left( x \right)$, which is completely arbitrary, we eliminate $a^0$ and $\Pi^0$ because they add nothing to the description of the system. 
%Note that the term containing $a_{0}$ can be absorbed by redefining the function $w(x)$. 
With this, the extended Hamiltonian is now given by 
\begin{eqnarray}
{H} &=& \int {{d^3}x} \left\{ { w(x)\left( {{\partial _i}{\Pi ^i} + \frac{1}{2}{\bf v} \cdot {\bf b} - {J^0}} \right) + \frac{1}{2}{C_1}{{\bf e}^2}} \right\}   \nonumber\\
 &+& \int {{d^3}x} \left\{ {\frac{1}{2}{D_2}\,{{\left( {{{\bf B}_B} \cdot {\bf e}} \right)}^2} + \frac{1}{2}{C_1}{{\bf b}^2} + {C_1}{{\bf B}_B} \cdot {\bf b}} \right\}, 
 \label{ener35}
\end{eqnarray}
where we have defined $w(x) \equiv u_1 (x) - a_0 (x)$.

According to usual procedure, we now proceed to introduce a gauge condition such that the full set of constraints become second class. Hence, we choose the gauge fixing condition as \cite{Gaete97}:
 \begin{equation}
\Gamma _2 \left( x \right) \equiv \int\limits_{C_{\zeta x} } {dz^\nu }
a_\nu\left( z \right) \equiv \int\limits_0^1 {d\lambda x^i } a_i \left( {
\lambda x } \right) = 0,  \label{ener40}
\end{equation}
where $\lambda$ $(0\leq \lambda\leq1)$ is the parameter describing the
space-like straight path $x^i = \zeta ^i + \lambda \left( {x - \zeta}
\right)^i $, and $\zeta $ is a fixed reference point. There is no essential loss of generality if we restrict our considerations to $\zeta^i=0$. One thus obtains the only non-vanishing equal-time Dirac bracket:
\begin{eqnarray}
\left\{ {a_i \left( {\bf x} \right),\Pi ^j \left( {\bf y} \right)} \right\}^ * &=& \delta{_i^j} \,\delta ^{\left( 3 \right)} \left( {{\bf x} - {\bf y}} \right) 
- \partial _i^x 
\int\limits_0^1 {d\lambda \,x^j } \delta ^{\left( 3 \right)} \left( {\lambda
{\bf x}- {\bf y}} \right).  \label{ener45}
\end{eqnarray}

With the aid of this result one can write the Dirac brackets in terms of the magnetic and electric fields as
\begin{equation}
{\left\{ {{b_i}\left( {\bf x} \right),{b_j}\left( {\bf y} \right)} \right\}^ * } = 0, \label{ener50-a}
\end{equation}
\begin{eqnarray}
{\left\{ {{e_i}\left( {\bf x} \right),{b_j}\left( {\bf y} \right)} \right\}^ * } &=& \frac{1}{{{C_1}}}{\varepsilon _{ijk}}{\partial ^k}{\delta ^{\left( 3 \right)}}\left( {{\bf x} - {\bf y}} \right) 
- \! \frac{{{D_2}}}{{C_1^2\det D}}{B_{Bi}}{B_{Br}}{\varepsilon _{rjk}}{\partial ^k}{\delta ^{\left( 3 \right)}}\left( {{\bf x} - {\bf y}} \right),
\label{ener50-b}
\end{eqnarray}
and
\begin{eqnarray}
{\left\{ {{e_i}\left( {\bf x} \right),{e_j}\left( {\bf y} \right)} \right\}^ * } &=& - \frac{1}{{C_1^2}}{\varepsilon _{ijk}}\,{v_k}\,{\delta ^{\left( 3 \right)}}\left( {{\bf x} - {\bf y}} \right) \nonumber\\
&+& \frac{{{D_2}}}{{C_1^3\det D}}\left( {{B_j}{B_q}{\varepsilon _{iqk}} - {B_i}{B_q}{\varepsilon _{jqk}}} \right) 
 {v_k}\,{\delta ^{\left( 3 \right)}}\left( {{\bf x} - {\bf y}} \right), \label{ener50-c}
\end{eqnarray}
where $det\,D = 1+\frac{D_2}{C_1}{\bf B}_B^2$.

We can now write the equations of motion for the electric and magnetic fields in the form
\begin{equation}
{\dot e_i}\left( x \right) =  - \frac{1}{{{C_1}}}\left( {1 - \frac{{{D_2}}}{{{C_1}}}{\bf B}_B^2} \right){\varepsilon _{ijk}}{v_k}{e_j} + \frac{1}{{\det D}}{\varepsilon _{ijk}}{\partial ^k}{b_j}, \label{ener55-a}
\end{equation}
and
\begin{equation}
{\dot b_i}\left( x \right) = {\varepsilon _{ijk}}{\partial _{j}}e_{k}. \label{ener55-b}
\end{equation}

Similarly, we write Gauss's law as
\begin{equation}
{C_1}\,\det D\,{\partial _i}{e^i} + {\bf b} \cdot {\bf v} - {J^0} = 0. \label{ener60}
\end{equation}
In this last line we used that the external magnetic field, ${{\bf B}_B}$, has a fixed direction in space.

We shall now consider the static case, that is, eqs. (\ref{ener55-a}) and (\ref{ener55-b}) must vanish. In this way one encounters 
\begin{equation}
e_{i} = \partial_{i}\Phi, \label{ener65}
\end{equation}
and
\begin{equation}
\Phi  = \frac{{{\nabla ^2}}}{{\left[ {{\nabla ^2}\left( {{C_1}\det D\,{\nabla ^2} + \xi \,{{\bf v}^2}} \right) - \xi\, {v^i}{\partial _i}{v^j}{\partial _j}} \right]}}\left( { - {J^0}} \right), \label{ener70}
\end{equation}
where $\xi  = \frac{1}{{{C_1}}}\det D\left( {1 - \frac{{{D_2}}}{{{C_1}}}{\bf B}_B^2} \right)$.

Incidentally, the above expression is analogous to that encountered in our previous work \cite{EPL}. In view of this situation, we skip all the technical details and refer to \cite{EPL} for them. Accordingly, the static potential between static $q$-charged point-like sources turns out to be 
\begin{eqnarray}
V &=&  - \frac{{{q^2}}}{{4\pi }}\frac{1}{{{C_1}\left( {1 + \frac{{{D_2}}}{{{C_1}}}{\bf B}_B^2} \right)}}\frac{1}{r} \nonumber\\
&+& \frac{{{q^2}}}{{4\pi }}\frac{1}{{{C_1}\left( {1 + \frac{{{D_2}}}{{{C_1}}}{\bf B}_B^2} \right)}}\frac{{{{\bf v}^2}}}{{C_1^2}}\left( {1 - \frac{{{D_2}}}{{{C_1}}}{\bf B}_B^2} \right)z\ln \left( {\frac{{z + r}}{{2z}}} \right), 
\label{ener75}
\end{eqnarray}
where $r = \sqrt {{{\bf x}^2} + {{\bf y}^2}}$.

Finally, by making use of equations  (\ref{NLCFJ80-d}), the statice potential profile can be rewritten in the form 
\begin{eqnarray}
V &=&  - \frac{{{q^2}}}{{4\pi }}\frac{{\left( {1 + {{{\bf B}_B^2} \mathord{\left/
 {\vphantom {{{\bf B}_B^2} {2{\beta ^2}}}} \right.
 \kern-\nulldelimiterspace} {2{\beta ^2}}}} \right)}}{{\left( {1 + {{{\bf B}_B^2} \mathord{\left/
 {\vphantom {{B_B^2} {{\beta ^2}}}} \right.
 \kern-\nulldelimiterspace} {{\beta ^2}}}} \right)}}\frac{1}{r} \nonumber\\
&+& \frac{{{q^2}}}{{4\pi }} \, {{\bf v}^2} \,   \frac{{{{\left( {1 + {{{\bf B}_B^2} \mathord{\left/
 {\vphantom {{B_B^2} {2{\beta ^2}}}} \right.
 \kern-\nulldelimiterspace} {2{\beta ^2}}}} \right)}^3}\left( {1 - {{{\bf B}_B^2} \mathord{\left/
 {\vphantom {{{\bf B}_B^2} {{\beta ^2}}}} \right.
 \kern-\nulldelimiterspace} {{\beta ^2}}}} \right)}}{{\left( {1 + {{{\bf B}_B^2} \mathord{\left/
 {\vphantom {{B_B^2} {{\beta ^2}}}} \right.
 \kern-\nulldelimiterspace} {{\beta ^2}}}} \right)}}\,\,z\,\ln \left( {\frac{{z + r}}{{2z}}} \right).
\end{eqnarray}
Of particular concern to us is the effect of the parameter that carry the LSV message on the static potential profile. To be more precise, the logarithmic correction to the usual Coulomb potential.

\section{Concluding Remarks}

In summary, we have explored the physical consequences on the vacuum exposed to non-linear electrodynamics coupled to parameters that signal violation Lorentz-symmetry. This new vacuum behaves as a medium with nontrivial refractive indices $n_ \bot$ and 
$n_ \parallel$. Hence, a charged particle interacting with this new medium can emit Cherenkov radiation if the velocity of the charged particle is larger the velocity of light in the medium.
In addition, we have calculated interaction potential for logarithmic electrodynamics with a term CFJ. Our calculations show that the interaction energy, at leading order in $\beta ^{2}$, is similar to our previous calculation \cite{EPL}.

Finally, having in mind that a number of astrophysical phenomena 
are strongly influenced by magnetic fields present in
the cosmo and that many of these phenomena can be systematically
described only in terms of Magnetohydrodynamics (MHD) 
processes \cite{Alfven, Poedts}, we are taking as a step forward
the re-assessment of the MHD equations in presence
of the space-time anisotropy parametrized by the CFJ $4$-vector,
$v_\mu$, along with non-linear effects. We recall the the MHD 
equations arise by coupling the equations of fluid mechanics
with Maxwell equations. New effects introduced in the latter
may modify the MHD set-up and induce corrections or even
unveil interesting effects in astrophysical plasmas, such as 
stellar coronae, magnetospheres, regions around pulsars/
magnetars/kilonovas,  black hole accretion disks and AGN jets.
More than $90 \%$ of visible matter are found in plasma state;
this justifies an endeavor of investigating MHD equations in
presence of post-Maxwellian terms induced by physics beyond 
the Standard Model, as the ones we have considered in this
contribution.

As a natural path open, particular wave modes that stem from the 
MHD equations, such as Alfvén waves and slow/fast magnetosonic waves,
should be re-assessed. These waves, that experience no dispersion, are 
expected to become dispersive by virtue of the LSV term in the Maxwell equations. 
Recognizing the relevance of MHD waves in astrophysical plasmas \cite{Swanson}, 
we intend to concentrate efforts to re-inspect the relevant phenomenon of 
magnetic reconnection by considering now the modified MHD equations.\\

{\bf Acknowledgments}: 

One of us (P. G.) was partially supported by ANID PIA / APOYO AFB220004.

\end{document}